\documentclass[%
 aip,
 amsmath,amssymb,
 reprint,%
longbibliography,
]{revtex4-2}

\pdfoutput=1 %for arXiv

\usepackage{graphicx}% Include figure files
\usepackage{dcolumn}% Align table columns on decimal point
\usepackage{bm}% bold math
\usepackage{mathtools}
\usepackage[utf8]{inputenc}
\usepackage[T1]{fontenc}
\usepackage{mathptmx}

\begin{document}

\newcommand{\mat}[1]{{\mbox{\boldmath{${\mathrm{#1}}$}}}}
\newcommand{\oper}[1]{\hat{\mbox{\boldmath{${\mathrm{#1}}$}}}}
\newcommand{\boltm}[1]{\mbox{\boldmath{$#1$}}}
\newcommand{\bra}[1]{\langle #1|}
\newcommand{\ket}[1]{|#1 \rangle}
\newcommand{\bracket}[3]{\langle#1\left|#2 \right|#3\rangle}
\newcommand{\braket}[2]{\langle#1 |#2\rangle}
\newcommand{\expectation}[1]{\langle #1 \rangle}
\newcommand{\creation}[1]{\hat{a}_{#1}^{\dagger}}
\newcommand{\annihilation}[1]{\hat{a}_{#1}}
\newcommand{\at}{$\hbar/{\textrm{E}_{\textrm{h}}}$}
\newcommand{\Eh}{${\textrm{E}_{\textrm{h}}}$}
\newcommand{\hb}{${\mathrm{\hbar}}$}
\newcommand{\bohr}{\,${\mathrm{a_0}}$}

\newcommand{\ba}{\begin{eqnarray}}
\newcommand{\ea}{\end{eqnarray}}
\newcommand{\bc}{\begin{center}}
\newcommand{\ec}{\end{center}}
\newcommand{\be}{\begin{equation}}
\newcommand{\ee}{\end{equation}}
\newcommand{\EA}{\emph{et al.}\xspace}
\newcommand{\ra}{\rightarrow}
\newcommand{\bi}{\begin{itemize}}
\newcommand{\ei}{\end{itemize}}
\newcommand{\rb}{{\bf r}}
\preprint{AIP/123-QED}

\title{A Quantum-compute Algorithm for Exact Laser-driven Electron Dynamics in Molecules}% Force line breaks with \\

\author{Fabian Langkabel}
 \altaffiliation[Also at the ]{Institute of Chemistry and Biochemistry, Freie Universit\"{a}t Berlin, Arnimallee 22, 14195 Berlin, Germany}%Lines break automatically or can be forced with \\
\author{Annika Bande}%
 \email{annika.bande@helmholtz-berlin.de}
 \homepage{https://hz-b.de/theochem}
\affiliation{ 
Helmholtz-Zentrum Berlin f\"ur Materialien und Energie GmbH, Hahn-Meitner-Platz 1, 14109 Berlin, Germany%\\This line break forced with \textbackslash\textbackslash
}%

\date{\today}% It is always \today, today,

\begin{abstract}
In this work, we investigate the capability of known quantum-computing algorithms for fault-tolerant quantum computing to simulate the laser-driven electron dynamics in small molecules such as lithium hydride. These computations are executed on a quantum-computer simulator. Results are compared with the time-dependent full configuration interaction method (TD-FCI). The actual wave packet propagation is closely reproduced using the Jordan-Wigner transformation and the Trotter product formula. In addition, the time-dependent dipole moment, as an example of a time-dependent expectation value, is calculated using the Hadamard test. In order to include non-Hermitian operators in the dynamics, a similar approach to the quantum imaginary time evolution (QITE) algorithm is employed to translate the propagator into quantum gates. Thus, ionization of a hydrogen molecule under the influence of a complex absorbing potential can be simulated accurately. All quantum computer algorithms used scale polynomially rather than exponentially as TD-FCI and therefore hold promise for substantial progress in the understanding of electron dynamics of increasingly large molecular systems in the future.
\end{abstract}

\maketitle

\section{Introduction}\label{SecIntro}
%Electron Dynamics: Why cool and relevant--------------------------------------------------
Ultrafast electron dynamics has become a vivid research field in the last couple of years due to the recent developments in experimental techniques for single-molecule and strong-field spectroscopy, which allow for the resolution both of sub-nanometer distances and attosecond times scales.\cite{Simoncig2017,Porat2018,Garcia2021} Accompanied by theoretical endeavors the interpretation of laser-driven high-harmonics generation\cite{Santra2006,White2015} or the formation of excitons\cite{Langkabel2022} becomes complete. Beyond these academic examples, electron dynamics also plays a vital role in chemistry, as molecular bond formation itself is involves substantial electron rearrangement. Furthermore, photo-induced processes are widely applied in chemicals production, i.e. in photo-catalysis, where the excitation of the electronic systems leads to charge transfer,\cite{Weber2020} bond breaking,\cite{Pino2010, Zhou2016, Pandeya2021} electron solvation,\cite{Maier2001,Buchner2022} etc. And indeed, electron dynamics does not end at this intermediately-sized systems, but is likewise important in nano-structured matter, like the process of human vision in the light-harvesting proteins,\cite{Thyrhaug2021} or the electronic decay in semiconductor quantum dots allowing for the generation of highly-brilliant single photons useful for quantum-information technologies.\cite{Reindl2019}%\cite{Bande2011, Langkabel2019, Pont2016} 

%Electron Dynamics: typical computation on classical computers---------------------------
As it is the case for many complex questions in natural sciences, their understanding - the foundation to utilize and design of technology - is impossible without theoretical modeling. This is even more true for quantum mechanics, including specifically quantum dynamics. There, one seeks to solve the time-dependent electronic Schr\"odinger equation using basically the same formulations as were developed some decades earlier for nuclear quantum dynamics.\cite{Elghobashi2003} Effectively, the Hamilton energy operators of the electronic system and those of external stimuli (driving laser fields or dissipating vibrations, some of them non-Hermitian) are applied to an initial electronic wave packet, e.g. for the system in its ground state, in successive times steps. Hence, the evolution is a time propagation leading to equations of motion. Established formalisms differ predominantly in the functional form of the wavefunction and thus affect for the amount of electron correlation accounted for. In this work we apply the widely-used time-dependent configuration interaction (TD-CI) method,\cite{Krause2005,Greenman2010, Sonk2011, Ulusoy2018,Langkabel2022}. For TD-CI the ``holy grail" is the full-CI (FCI) variant, allowing all possible electron configurations. TD-FCI, however, scales exponentially with the number of electrons and is therefore beyond reach for all but the smallest molecules. Nowadays, for larger molecules or nanocrystals\cite{Weber2020, Krause2021}, TD-CIS computations, including only singly-excited configurations, are the upper limit as determined by conventional compute resources, as it scales only polynomially.

% Use Quantum Computers: Motivation via Quantum Mechanics, History ---------------------------
Already from the birth hour of quantum mechanics in the 1920s it was recognized that most quantum-mechanical problems would remain unsolved given their sheer number of interacting particles incorporated in the above-mentioned wavefunctions. With the advent computers, quantum-mechanical problems were limited by numerical methods. The idea to use quantum-mechanical hardware -- quantum computers -- for such calculations was later formulated by Feynman.\cite{Feynman1982} Their inherent advantage is that every quantum bit (qubit) of information is not only ``0" or ``1" as traditional bits, but any complex value within those limit. Quantum registers (fully) correlate the individual qubits, which means a complete description of interactions among the quantum-mechanical particles. Every operator of the Hamiltonian would translate into a quantum gate operating on the qubits defining the molecular state.    

% Use Quantum Computers: Motivation via Construction of Hardware
Besides the obvious theoretical advantages, it took nearly a century to initiate executing quantum-mechanics computations on relevant problems by quantum computers, and most progress goes back to some eruptive advances in recent years. On the one hand, quantum hardware was build. The most prominent realization are superconducting qubits operated through resonators as gates, because they are already on the market with currently over hundred qubits.\cite{Ball2021} However, it is still unclear whether superconducting qubits will continue to lead in terms of error rates, scalability, and the required ultracold environment, given the rapid experimental progress in other areas of physics and engineering driving the development of alternative architectures, e.g. those based on ions controlled by traps, photons controlled by interferometers, electronic states in cold atoms controlled by optical lasers, spins controlled by magnetic fields, etc. All developments target the construction of an error-free universal quantum computer,\cite{DiVincenzo2000} which operates on well-defined and temporary stable qubits whose registers are scalable.

%Use Quantum Computers: Motivation via Algorithms: True vs. Simulator
On the other hand, algorithms for quantum computers have been already developed for application in various fields, even outside of physics and chemistry. They are divided according to whether they can already be executed on presently existing error-prone NISQ computers (noisy intermediate-scale quantum computers), or whether they are designed for an error-free quantum computer, such that currently they are executed on quantum computer simulators.

% Use Quantum Computers: Motivation via other Chemistry applications
Given the direct correspondence of a qubit to a single-particle state function in a wavefunction for the electrons in a molecule, several algorithms have already been developed for quantum chemistry and materials calculations. As electronic structure computations, which solve the time-independent electronic Schr\"odinger equation to obtain properties like molecular structure, their energy spectrum, and orbitals, were developed over decades along with high-performance computer technology, it is no surprise to see already the first quantum algorithms in this field.\cite{Cao2019, McArdle2020} These algorithms include, for example, the Jordan-Wigner transformation for the representation of the chemical Fock space to the Hilbert space of the qbits\cite{Jordan1928} and numerous algorithms to solve the eigenvalue problem posed in the time-independent electronic Schr\"odinger equation, giving energy eigenvalues of several states of the molecule. A prominent one is the quantum phase estimation (QPE)\cite{Kitaev1995, Abrams1999, MichaelA.Nielsen2010} in which a central part is the Hamiltonian simulation, the short-time propagation of the electronic system using a Li-Trotter-Suzuki decomposed propagator\cite{Trotter1959, Suzuki1976} followed by the quantum Fourier transformation to obtain the phases and the state energies. The lowest excited states of water or lithium hydride, for example, have been computed at FCI level of theory.\cite{AspuruGuzik2005} Related to the QPE is the Hadamard test,\cite{Aharonov2008} which also uses Hamiltonian simulation to determine expectation values of arbitrary operators. Another interesting algorithm is the quantum imaginary time evolution (QITE), which offers the possibility to obtain eigenstate function (in addition to the eigenvalue) on a quantum computer which can be applied in follow-up quantum algorithms. The theoretical formulation of the algorithms as well as the successful execution demonstrate an embarrassing advantage in terms of calculation time over traditional computations: instead of an exponential scaling for FCI on traditional computers, its quantum computation results in a polynomial scaling only.

This work seeks to extend the application of quantum algorithms in the domain of error-free quantum computers for the simulation of electron dynamics. The target are TD-FCI calculations for the light-induced dynamics in small molecules like H$_2$ and LiH in order to benchmark two- and four-electrons systems. This requires the representation of wavefunction and operators in the quantum computer, adopting the Jordan-Wigner encoding.\cite{Jordan1928} The next ingredient is encoding of the time-propagation series, a step-wise application of operators to an initial function, which is realized using the Li-Trotter-Suzuki decomposition\cite{Trotter1959, Suzuki1976} for the propagator, and has already been performed for dynamics with so-called 1-sparse Hamiltonians.\cite{Wiebe2011} This allows to compute excitation according to the extremes of laser pulse scenarios, namely state-to-state transitions achieved with a weak, long, and resonant pulse as well as creation of a wave packet with many states populated using intense short pulses. Additionally, the ionization process in H$_2$ was investigated, which results in a free electron leaving the molecule. In order to represent the continuum, a complex absorbing potential (CAP) in energy space is used in form of an additional operator in the Hamiltonian.\cite{Klinkusch2009, Coccia2017} This operator, however, poses a particular challenge to quantum computing, because it is non-Hermitian, i.e., it leads to non-unitary transformations during the propagation that cannot be translated into quantum gates. We realize its implementation for the dynamics via a QITE framework.\cite{Motta2019}. Finally, every electron-dynamics calculation is evaluated by routines for the time evolution of properties, e.g., the changing dipole moment through electronic rearrangement. Here, we employ the Hadamard test\cite{Aharonov2008} to obtain their respective time-dependent expectation values (at least for Hermitian operators). 

%Specific details on paper organization and Programs
The paper is structured such that in the theory section \ref{SecTheo} the TD-CI method is reviewed  (Sec. \ref{SecTheo_TDCI}) followed by the specified introduction of the above mentioned quantum algorithms for initial state and operator representation (Sec. \ref{SecTheo_QComp_OpWf}), propagation (Sec. \ref{SecTheo_QComp_Prop}), and analysis (Sec. \ref{SecTheo_QComp_Ana}). Special emphasis lies in the implementation of non-Hermitian operators for propagation (Sec. \ref{SecTheo_nonHerm}). In the computational details (Sec. \ref{SecComp}) the group's own program Jellyfish  - capable of traditional and quantum algorithms - for TD-CI propagations is briefly sketched. Details on the chemistry and laser are heading the result subsections for the symmetric H$_2$ molecule - the work horse of developments in electron dynamics - are given in Sec. \ref{SecRes_H2St2St}, for the asymmetric LiH in Sec. \ref{SecRes_LiH}, and the ionization dynamics of H$_2$ in Sec. \ref{SecRes_H2Ion}. In the conclusive discussion (Sec. \ref{SecConclusion}) we critically evaluate the performance of the chosen quantum algorithms, name alternatives and sketch necessary steps to advance the novel field of quantum-computed electron dynamics towards a useful application of quantum computers giving insight into chemical processes, that cannot computed with any existing traditional computer.

\section{Theory}\label{SecTheo}
\subsection{Time-dependent Configuration Interaction}\label{SecTheo_TDCI}
The time-dependent electronic Schr\"odinger equation (TDSE),
\begin{equation}\label{tdse}
  i \frac{\partial  \ket{\Psi_{el}(t)}}{\partial t}=\left[\hat{H}_{el} + \hat{\vec{\mu}}\cdot\vec{F}(t)\right]\ket{\Psi_{el}(t)} \quad,
\end{equation}
is the core of every many-electron dynamics calculation. It describes the evolution of an electronic wavefunction  $\ket{\Psi_{el}(t)}$ with time $t$ from an initial state at $t_0 = 0$. The electronic Hamiltonian, $\hat{H}_{el}=\hat{T}_{el} + \hat{V}_{el,nuc} +\hat{V}_{el,el}$, composed of the kinetic energy operator of electrons, their binding to the nuclei, and their Coulomb interaction determines the eigenstates of the molecule, whereas the laser, $\vec{F}(t)$, induces the dynamics and makes the system explicitly time-dependent. A typical expression for a laser pulse is 
\begin{equation}\label{laser}
{\vec{F}(t)=\vec{f}_0 \sin(\omega (t-t_p)) \cdot \cos^2\left(\frac{\pi}{2\sigma} (t-t_p) \right)}
\end{equation}
and is incorporated in Eq.\,\ref{tdse} via the semi-classical dipole approximation. This linearly-polarized oscillating external field with carrier frequency $\omega$ is applied as a $\cos^2$ pulse for a duration determined by its width $\sigma$ centered at $t_p$. Its coupling strength to the electronic system depends on the molecular dipole operator $\oper{\vec{\mu}}=-\sum_i^N \vec{r}_i + \sum_A^{N_A} Z_A {\vec{R}}_A$ accounting for $N$ electrons and $N_A$ nuclei at the respective positions $\vec{r}_i$ and $\vec{R}_A$ and carrying charges of unity (in atomic units) and $Z_A$, respectively. All dynamical calculations are carried out for a fixed molecular geometry.

Several options exist for the functional representation of the wavefunction, one of which consolidates the TD-CI method. There, the wavefunction is given in the basis of configuration interaction (CI) eigenfunctions $\Phi_i$ which are, depending on the desired level of accuracy, composed of singly, doubly, etc. configurations. One configuration is a normalized, antisymmetrized products of spin orbitals which are also called Slater determinants. The time-dependent wavefunction $\Psi_{\rm el}(t)$ in TD-CI is represented as linear combination of eigenstates with the time-dependent coefficients $B_i(t)$, so that 
\begin{equation}\label{td-ci-expans}
  \ket{\Psi_{\rm el}(t)} = \sum\limits_{i} B_{i}(t)\ket{\Phi_i} \quad.
\end{equation}
Solving the TDSE (Eq.\,(\ref{tdse})) in the basis of Eq.\,(\ref{td-ci-expans}) leads to equations of motion for the individual coefficients being varied in the mean field of the others. Its formal solution 
\be 
B_i(t)= e^{-i \hat{H} t} B_i(0) \label{EQ_Propagation} \quad,
\ee
contains the exponential propagator term, which is the reason that time evolutions are often also called propagations. It acts on the initial coefficient $B_i(0)$ to obtain the coefficient $B_i(t)$ at later times.
For the numerical solution on traditional computers, the propagation is executed in a series of very small time steps $\Delta t$ (short-time approximation). This is a special case to the later introduced Trotter decomposition of propagators (cf. Sec. \ref{SecTheo_QComp_Prop}).\cite{Trotter1959} 

The results from a TD-CI calculation can be analyzed in different ways, straightforwardly, by time-dependent populations of the CI states as 
\begin{equation}\label{EQ_Population}
P_i(t) = |\langle \Phi_i| \Psi(t) \rangle|^2 = |B_i(t)|^2 \quad, 
\end{equation}
or by evaluation of various expectation values like the position of the wave packet in space or the dipole moment
\begin{equation}\label{EQ_Dipole}
\mu(t) = \langle \Psi(t) |\hat{\mu}| \Psi(t) \rangle \quad,
\end{equation}
in order to follow the charge displacement during the dynamics. Utilising the expectation value of the density operator $\hat{\rho}(r)$, density visualization techniques exist,\cite{Langkabel2022} that can show the difference density among the current and the initial state, 
\begin{equation}\label{DensityDifference}
\begin{multlined}
\Delta_0(t) = \rho (r,t) - \rho (r,0) = \left\langle \Psi(t) | \hat{\rho}(r) | \Psi(t) \right\rangle \\ - \left\langle \Psi(0) | \hat{\rho}(r) | \Psi(0) \right\rangle \quad.
\end{multlined}
\end{equation}

%\subsection{Complex-Absorption Potentials in TDCI}
Electron dynamics may lead to ionization where an electron leaves the molecule. This can either happen through direct processes induced by intense or high-frequency pulses or through indirect processes such as auto-ionization\cite{Aberg2012} or inter-particle Coulombic decay.\cite{Bande2011, Langkabel2019} Here, it is of interest to determine the degree of ionization. This can be done via the wavefunction overlap with a given complex absorbing potential (CAP). Various CAPs exist. For CI states in an atom centered basis, a heuristic model is applicable.\cite{Klinkusch2009, Coccia2017}
Therein, an inverse lifetime (rate $\gamma_p$) is added in form of a negative imaginary contribution to all positive orbital energies ($\epsilon_p > 0$) included in the Slater determinants, i.e., $\epsilon_p - i \gamma_p/2$. It is defined through  $\gamma_p = \left( d \sqrt{2 \epsilon_p} \right)^{-1}$, where $d$ is a characteristic escape length relating to the kinetic energy of the outgoing electron.
Formally, the respective CAP Hamiltonian in second quantization reads
\begin{equation}\label{Eq_CAPDefinition}
  \hat{V}_{CAP} = - \frac{i}{2} \sum_p \gamma_p \creation{p} \annihilation{p} \quad ,
\end{equation}
where $\creation{p}$ and $\annihilation{p}$ are the creation and annihilation operator and $\gamma_p = 0$ for $\epsilon_p \leq 0$. Adding such a CAP operator leads to adding an inverse lifetime $\Gamma_n$ to the CI state energies as well
\begin{equation}\label{}
E_n^{CI} \rightarrow E_n^{CI} - \frac{i}{2} \Gamma_n \quad.
\end{equation}

\subsection{Quantum Computing Algorithms}\label{SecTheo_QComp}
\begin{figure}
\includegraphics[width=0.48\textwidth,clip=]{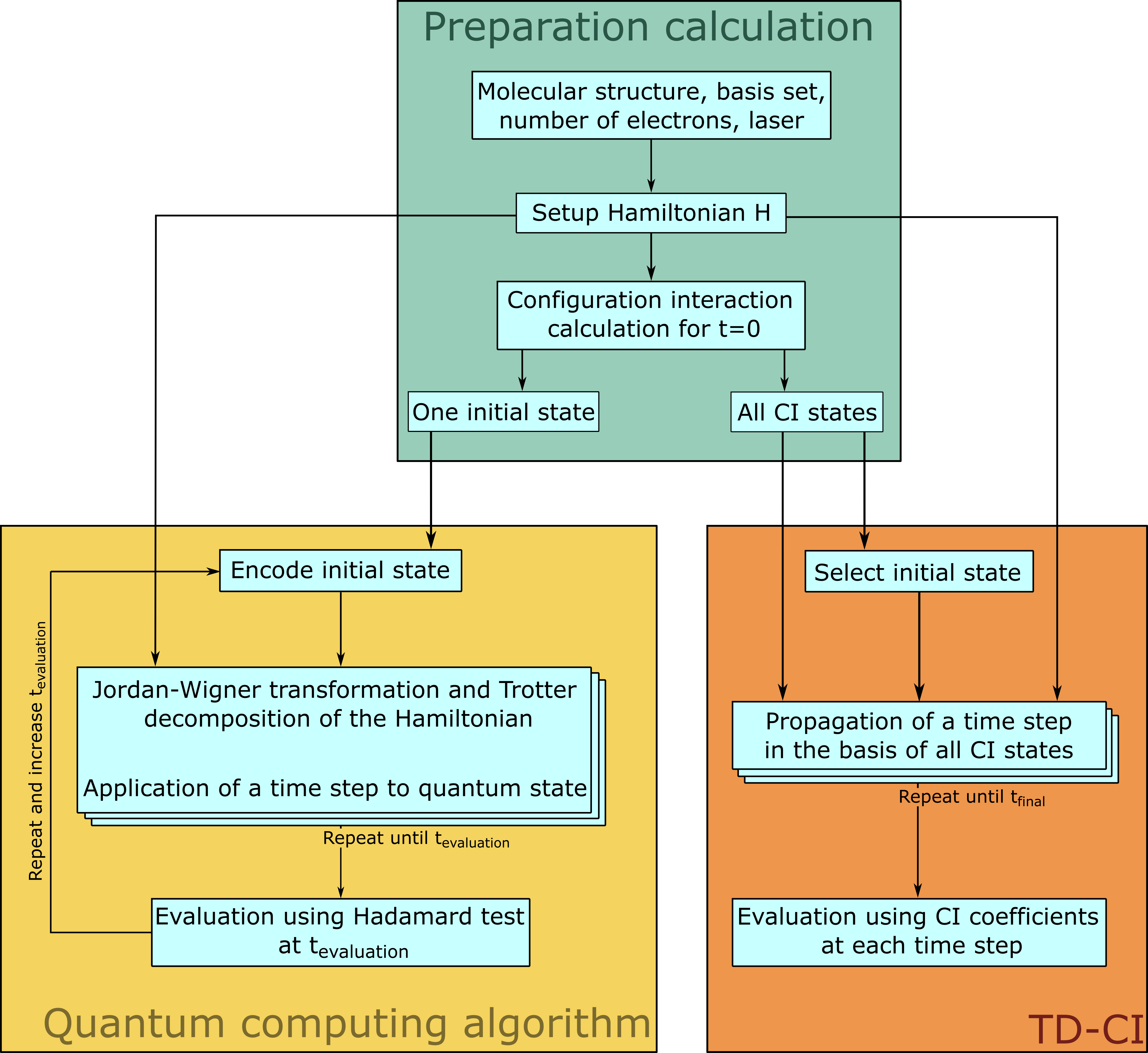}
\caption{Design of the algorithms used to perform and evaluate electron dynamics calculations using traditional and quantum computers.\label{fig:flowchart}}
\end{figure}

Performing such dynamics simulations on a quantum computer is not a one-to-one translation of algorithms for traditional computer architectures to qubit-based ones. Both algorithms have the same procedure consisting of generating an initial wavefunction, propagating the wavefunction, and evaluating the propagation, however, all the individual steps between the algorithms differ fundamentally from each another. Fig. \ref{fig:flowchart} represents a flow chart for ``traditional" TD-CI algorithm (orange box) and the procedure for quantum computing (yellow box), including the respective algorithms employed. Both algorithms can be initiated after the same preparative calculation (green box) executed with a time-independent method and here depicted without recognition of the underlying computer systems. In TD-CI an arbitrary state vector can be selected as initial state and all CI states are needed, while for the quantum computer algorithm only one initial state vector has to be calculated and Jordan-Wigner transformed. Structurally, both algorithms then sequentially apply propagators for individual time steps on this initial state. In the quantum algorithm, these propagators are beforehand successively Jordan-Wigner transformed, Trotter decomposed, and translated into quantum gates. Evaluation completes the computational flow. While in the TD-CI algorithm the evaluation is usually done after the propagation, in quantum computing it is necessary to do this during the propagation. For that, the propagation is repeated several times until at the current time step $t_{evaluation}$, the Hadamard test is performed. For a completed propagation, the number of time steps $t_{evaluation}$ is increased until finally $t_{final}$ is reached. Details on all mentioned algorithms are described in the following sections.

%Initial States
\subsubsection{Jordan-Wigner encoding of States and Operators} \label{SecTheo_QComp_OpWf}
First of all, the initial wavefunction and the Hamiltonian included in the propagator need to be transferred to the quantum computer. This means, a system of indistinguishable fermions (Fock space) must be mapped to a system of distinguishable qubits (Hilbert space), which can be done in second quantization using the Jordan-Wigner (JW) encoding.\cite{Jordan1928} Here, the occupations of $M$ spin orbitals of a wavefunction are stored as $\ket{0}$ or $\ket{1}$ states of $M$ qubits. 

The quantum register is a superposition of $2^M$ basis states whereby in the JW encoding each of these basis states corresponds to a Slater determinant. For the initial state, any Slater det, such as the HF GS, can be used, but also any excited states generated by CI \cite{Wang2009, Babbush2015} as well as from multi-configurational self-consistent field methods.\cite{Sugisaki2018}

In order to translate time-independent and time-dependent Hamiltonians, including the CAP operator as in Eq.\,(\ref{Eq_CAPDefinition}), the creation $\creation{p}$ and annhilation operators $\annihilation{p}$, which increase or decrease the occupation of a spin orbital by 1 and thereby introduce a phase factor, have to be translated.
In JW encoding they are translated as
\begin{equation}\label{}
\begin{aligned}
\creation{p} = \hat{Q}^\dagger_p \otimes \hat{Z}_{p-1} \otimes ... \otimes \hat{Z}_0 \quad, \\
\annihilation{p} = \hat{Q}_p \otimes \hat{Z}_{p-1} \otimes ... \otimes \hat{Z}_0 \quad,
\end{aligned}
\end{equation}
where $\hat{Q}^\dagger_p = \frac{1}{2}(\hat{X_p} + i\hat{Y_p})$ and $\hat{Q}_p = \frac{1}{2}(\hat{X_p} - i\hat{Y_p})$ reflect the change of the spin orbital occupation. $\hat{X}_p$, $\hat{Y}_p$, and $\hat{Z}_p$ are the Pauli quantum gates/operators $\sigma_p$ acting on the $p$-th qubit. The string of $\hat{Z}$ operators reflects the phase vector, which is also called ``calculation of the parity of a state". The fermionic Hamiltonian can now be represented by a linear combination of products of single-qubit Pauli operators as
\begin{equation}\label{Eq_OperatorDecomposition}
\hat{H} = \sum_j \hat{h}_j = \sum_j h_j \prod_i \sigma_{i,j} \quad,
\end{equation}
where each term $\hat{h}_j$ acts locally on only one particle, such as the kinetic energy of an electron, or a small subset of particles, such as the interaction between two electrons. The number of individual terms scales polynomially with the number of particles.

While in JW encoding the occupation of the spin-orbitals is stored locally, the parity is stored non-locally. There are also more complicated approaches of encoding (Parity and Bravyi-Kitaev encoding \cite{Bravyi2002}) where the parity is stored locally, which in many cases leads to shortened quantum gate sequences. \cite{Seeley2012, Tranter2018}

\subsubsection{Hamiltonian Simulation for Real-Time Propagation} \label{SecTheo_QComp_Prop}
After mapping initial wavefunction and general operator expressions to the quantum computer, the qubit wavefunction is propagated in time by transcribing, e.g., the Li-Trotter-Suzuki decomposed propagator (cf. Eq.\,(\ref{EQ_Trotter})) into a series of quantum gates for each time step.\cite{Trotter1959,Suzuki1976, Berry2006} The propagator with the total Hamiltonian is thereby decomposed into a product of propagators with individual Hamiltonian terms $\hat{h}_j$ as in Eq.\,(\ref{Eq_OperatorDecomposition}), which can subsequently be translated into quantum gates. The decomposition can be performed in different orders, \cite{Suzuki1976, Berry2006} with the first one being
\begin{equation}\label{EQ_Trotter}
e^{-i\hat{H} \Delta t} = e^{-i \sum_j \hat{h}_j \Delta t} \approx \prod_j e^{-i\hat{h}_j \Delta t} + O(\Delta t^2) \quad,
\end{equation}
and the general $N$th order form being
\begin{equation}\label{EQ_Trotter2}
e^{-i\hat{H} \Delta t} = e^{-i \sum_j \hat{h}_j \Delta t} \approx \left( \prod_j e^{-i\hat{h}_j \Delta t/N} \right)^N \quad.
\end{equation}
Simulations of quantum dynamics including time-dependent operators using Trotter decomposition have already been successfully performed on quantum computers.\cite{Wiebe2011} More efficient methods like quantum walk, \cite{Berry2015} multiproduct formulae,\cite{Childs2012, Low2019} Taylor series expansion,\cite{Berry2009, Berry2015a} or qubitization \cite{Low2016, Low2016a, Low2017} algorithms have also been developed in the context of quantum time propagations. Note, that all the above mentioned real-time propagations rely on Hermitian operators and hence do not apply to open quantum systems. A respective extension is introduced in Sec. \ref{SecTheo_nonHerm}.

\subsubsection{Analysis via Hadamard Test} \label{SecTheo_QComp_Ana}
The last step of any quantum dynamics calculation is the analysis of the evolving wave packet through its time-dependent expectation values (see Sec. \ref{SecTheo_TDCI} and Fig. \ref{fig:flowchart}). After a quantum simulation this can be done with the Hadamard test. It is a variant of the quantum phase estimation algorithm, which was introduced in 2008 in the mathematical sciences to efficiently approximate the Jones polynomial using a quantum computer.\cite{Aharonov2008} Presently, it is often used as a subroutine of larger algorithms.\cite{Mitarai2018} 

Within the Hadamard test, the expectation value of any unitary operator $\hat{U}$ can be determined using an ancilla qubit and the controlled quantum gate to the unitary operator. By selection of the real or imaginary part of the expectation value $\langle \Psi |\hat{U}| \Psi \rangle$ that corresponds to the expectation value of the ancilla qubit in the given computational basis, it is determined via repeated execution of the algorithm until a accuracy of $\epsilon$ is reached in $O(1/ \epsilon)$ time. The Hadamard test is an indirect approach that is not destroying the quantum state. Thus it allows to use the state $(I \pm U)|\Psi \rangle \sqrt{2}$ after execution of the algorithm which is used for example in iterative quantum phase estimation.\cite{Dobsicek2007}

In this work, the Hadamard test serves for determination of the dipole moment $\mu$. Therefore, a unitary operator $e^{-i \mu \Delta x}$ is set up, where $\Delta x$ (in units of $1/ea_0$) is chosen in a way to minimalize the error of the respective Trotter decomposition. For the real dipole moment, it is sufficient to either determine the real or the imaginary part of the expectation value of its unitary operator. From that the expectation value of the dipole moment can be obtained by reformulation of the operator.

\subsection{Quantum Imaginary Time Evolution of non-Hermitian Operators}\label{SecTheo_nonHerm}
A Hamiltonian simulation as described in Sec. \ref{SecTheo_QComp_Prop} but executed for an open quantum system with a non-Hermitian operator cannot directly be translated into a unitary transformation due to a change of norm and thus cannot be expressed by a series of quantum gates. Nevertheless, a propagation with such operator is possible, e.g., with dilation methods,\cite{Sweke2015, Sweke2016, Sparrow2018, Hu2020, HeadMarsden2021, Hu2021} time-dependent variational methods,\cite{Endo2020} or the here implemented QITE algorithm.\cite{McArdle2019, Motta2019, Gomes2020}

The idea behind it is to exchange all those Trotter terms $e^{-i\hat{h}_j \Delta t}$ of Eq.\,(\ref{EQ_Trotter2}), which contain non-Hermitian operators $\hat{h}_j$, by unitary propagator terms $e^{-i \hat{A}^{m} \Delta t}$, under the condition that the new propagator leads to the same wavefunction but renormalizes it at the same time
\be
| \bar{\Psi}(t + \Delta t) \rangle = c_j e^{-i\hat{h}_j \Delta t} | \Psi(t) \rangle \approx e^{-i \hat{A}^{m} \Delta t} | \Psi(t) \rangle \quad.\label{EQ_QITEtarget}
\ee
The change in the norm $c^2$ must be determined to set up $\hat{A}^{m}$ and is obtained by measuring the expectation value of $\hat{h}_j$,
\be
c_j^2 = \langle \Psi | e^{-2 i \Delta t \hat{h}_j} | \Psi \rangle
\approx 1 - 2i \Delta t \langle \Psi | \hat{h}_j | \Psi \rangle \quad
\ee
and $c^2 = \sum_j c_j^2$.
While $\hat{h}_j$ acts locally on only a few qubits, correlations ensure that $\hat{A}^{m}$ may even act on several more, or all qubits $m$. $\hat{A}^{m}$ is determined by the topography of these $m$ qubits and is represented in the basis of all possible Pauli strings $\hat{\sigma}_I$ that can act on the $m$ qubits
\be
\hat{A}^{m} \equiv \sum_{I}a^{m}_I \hat{\sigma}_I \quad.
\ee
The equations to determine $\hat{A}^{m}$ can be taken from Motta et al. \cite{Motta2019}

\section{Computational Details}\label{SecComp}
All calculations on the molecules H$_2$ and LiH are performed with the groups owned program \emph{Jellyfish}. Electronic structure calculations executed on traditional hardware are preformed at FCI level to obtain the initial state and overall singlet-state basis for the TD-FCI dynamics calculations. The latter are executed using both, traditional and quantum algorithms for comparison. 

For the hydrogen molecule, the geometry is fixed at an equilibrium bond length of $1.40$ a.u.,\cite{Huber1979} and for lithium hydride this is $3.01$ a.u.\cite{Lovas2002} Both molecules are aligned in $z$ direction by convention. The FCI states and their energies are obtained by single-point calculations for a minimal STO-3G basis, returning four states for the two electrons of H$_2$ and 224 for LiH's four electrons.

Different electron dynamics calculations are executed for the two molecules. For H$_2$ a state-to-state transition to the first excited state is computed (Sec. \ref{SecRes_H2St2St}). The target state here is dominated by a transition from the highest occupied to the lowest unoccupied molecular orbital (HOMO-LUMO transition) with an excitation energy of $0.9673$ a.u., determining the laser frequency, and a relatively large transition dipole moment of $\vec{\mu}_{0,1;z} = 1.16$ a.u. oriented along the bond axis, which determines the laser strength and polarization. The duration of the $\pi$ pulse was fixed at $\sigma = 250$ a.u., hence, the field amplitude $\vec{f}_0 = 0.0108$ a.u. results directly from the condition $\sigma\cdot \vec{f}_{0}=\pi/\vec{\mu}_{0,1}$ for $\cos^2$ shaped $\pi$ pulses. 

Since the excited state is already in the continuum -- due to the minimal basis set -- the same dynamics with the possibility of ionization was performed. For this purpose, an additional non-Hermitian CAP operator is introduced (Eq. (\ref{Eq_CAPDefinition})), to quantify the amount of ionization through the loss of wavefunction norm. The characteristic escape length is set to value of $d=50$ a.u., as determined for the hydrogen atom by Coccia et al.\cite{Coccia2017} To counteract numerical instabilities when solving the necessary linear systems of equations for the QITE algorithm, a $\delta = 0.1$ was added as a regualrization on the main diagonal for the coefficient matrices.

For a state-to-state transition in LiH (Sec.\ref{SecRes_LiH}), the second excited singlet state was selected. This target state has a significantly different permanent dipole moment ($-1.816$ a.u.) than the ground state ($0.123$ a.u.), resulting in a charge-transfer excitations. It has an excitation energy of $0.183$ a.u., a transition dipole moment of $\vec{\mu}_{0,2;x} = -1.421$ a.u. and can be excited by a $x$-polarized laser. Note, that a degenerate  third singlet state with same transition dipole moment but $y$-polarization exists due to the molecular symmetry, which is not considered here. At same pulse duration as for H$_2$, the excitation is done with a relatively weak laser pulse of $f_{0}= 0.00885$ a.u. Complementary dynamics is performed with a shorter and more intense laser pulse in order to test the performance of a quantum computation for the creation of a wave packet, i.e., a superposition of multiple excited states. The laser is $z$-polarized and thus aligned with the transition dipole moment of $\vec{\mu}_{0,5;z} =-0.480$ a.u. of the fifths excited singlet state in $z$ direction. Under $\pi$ pulse conditions with an excitation frequency of $0.558$ a.u. and a width of $\sigma = 50$ a.u., a field strength of $f_{0}= 0.131$ a.u. results. The propagation was terminated at time $t= 125$ a.u. to obtain some of the oscillations of the wave packet created.

The TD-FCI calculations (cf. Sec. \ref{SecTheo_TDCI}) are done using an operator splitting technique.\cite{Bandrauka} A time step size of $1$ a.u. was used for all dynamics except the fast LiH stong-field process, where it was set to $0.1$ a.u.

The respective quantum algorithms (cf. Sec. \ref{SecTheo_QComp} and \ref{SecTheo_nonHerm}) are executed on classical computers as well. In order to simulate an error-free quantum computer, the library QuEST was integrated into the program \emph{Jellyfish}.\cite{Jones2019} Using QuEST rather than a true quantum computer has the advantage that the state of the quantum system can be read out, cached, and restored at any time without an additional simulation of a respective measurements, which would destroy the quantum state. Underlying the dynamics is the short-time approximation, where an optimal time step was sought within [$0.01;1$] a.u. Furthermore, the Trotter decomposition at first and second order is compared. Optimal settings are identified as second-order Trotter formula with a time step of 0.2 a.u. for weak laser pulses and a time step of 0.1 a.u. for the strong laser pulse through comparison of time-dependent state populations $P_i(t)$ (Eq. (\ref{EQ_Population})). The populations are read directly from the state vector of the quantum computer simulator and serve as a proof that the dynamics correspond to the dynamics with TD-CI. If the populations are to be determined on a real quantum computer, the Hadamard test can be used, which was utilized here to determine the time-dependent dipole moment. For this purpose, the dynamics was repeatedly simulated up to the corresponding time step and the Hadamard test was then repeated system-dependently several tens of thousands of times. To balance computational time and accuracy, the first-order Trotter decomposition was used for the Hadamard test. Further details on the algorithms are presented as results in the next section.

\section{Results}\label{SecResults}

\subsection{State-to-state Transition Dynamics of H$_2$}\label{SecRes_H2St2St}

\begin{figure}
\includegraphics[width=0.48\textwidth,clip=]{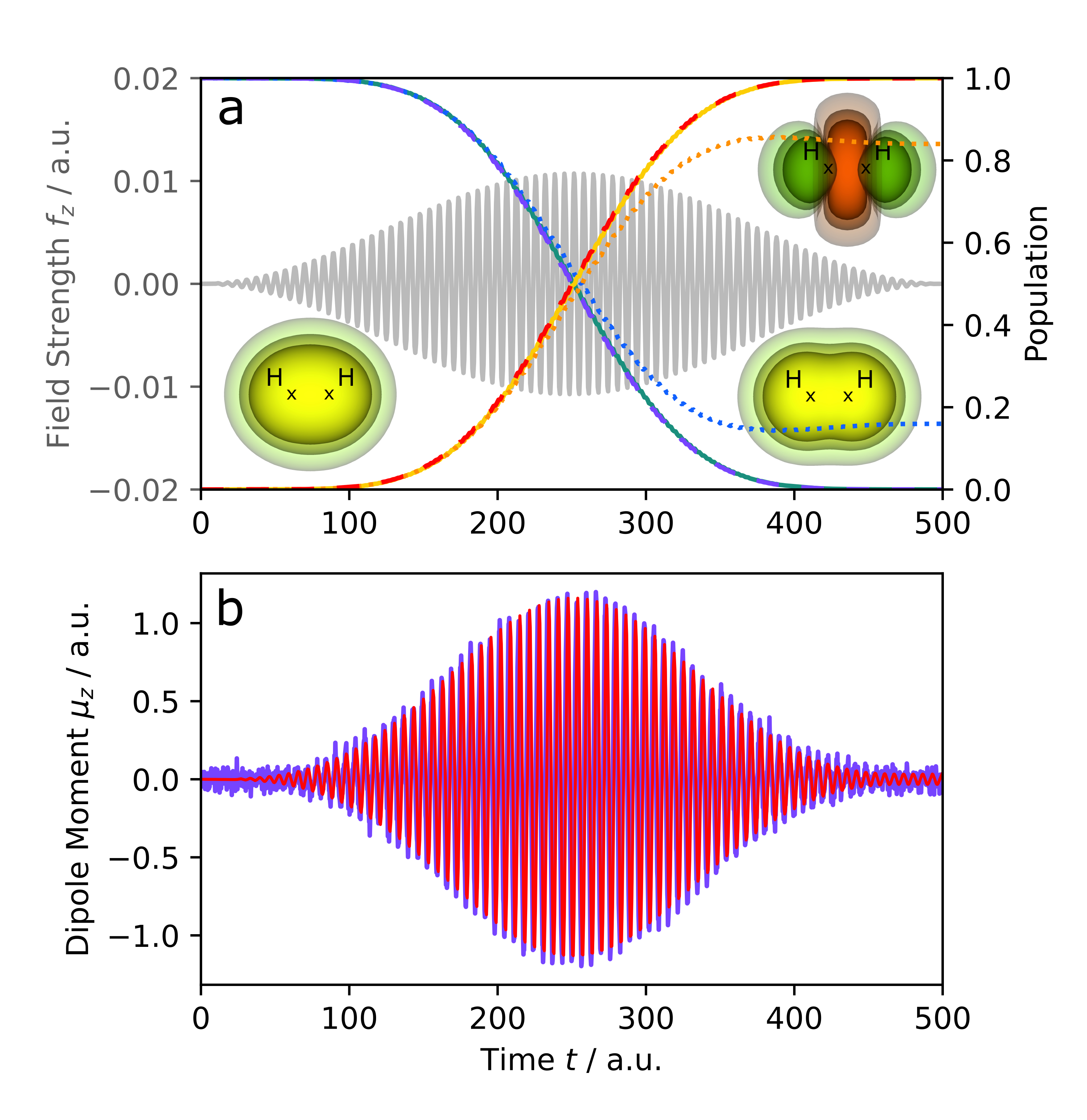}
\caption{\label{fig:h2} (a) Comparison of state population $P_i(t)$ during resonant $\pi$ pulse excitation (gray) in the hydrogen molecule obtained with TD-FCI (solid) and quantum computational dynamics, the latter with Trotter time steps of $1$ a.u. (dotted) and $0.2$ a.u (dashed). Insets show the electron density at the beginning and end of propagation as well as the difference density at the end. (b) Time-dependent dipole moment with TD-FCI (red) and Hadamard test (purple) with Trotter time steps of $0.2$ a.u.}
\end{figure}

The state-to-state transition dynamics of H$_2$ is summarized in Fig. \ref{fig:h2} (a) for both the quantum and the traditional algorithm. The driving laser pulse is depicted by light gray line in the background. All other lines depict the populations $P_i(t)$ of the ground and the target CI states. Here, the decreasing lines of cold colors (blue/green) correspond to the initial ground state, the electron density $\rho$ of which is shown on the bottom left, while the increasing lines of warm colors (yellow/red) correspond to the target first excited state. The difference density between the superimposed state at the last time step (actually exclusively the first excited one) and the initial one is shown on the right side, where red (green) signifies an electron density decrease (increase). The graphs show clearly the intended population inversion among ground and first excited state, the densities reveal that electron density is shifted from the center of the bond to the outside of the hydrogen atoms, thus destabilizing the molecule in the excited state, which agrees with previous studies on the H$_2$ excitation dynamics.\cite{Krause2007, Langkabel2022} Note, due to the minimal basis only four electronic states are included in the propagations, hence, a much shorter duration for the $\pi$ pulse could be chosen to obtain the population inversion. 

The effect of the time step size is also considered. For the quantum simulations two Trotter time steps are shown in Fig. \ref{fig:h2} and compared to the conventional computations which serves as a reference. Solid lines in lighter color stands for the traditional TD-FCI calculation (Sec. \ref{SecTheo_TDCI}) executed with an already optimized step size of $1$ a.u., which serves as reference. The broken lines in gradually darker color correspond to quantum simulations (cf. Sec. \ref{SecTheo_QComp}) executed with different time steps. The blue and orange dotted lines refer to time steps of $1$ a.u., identical to those for standard TD-CI, in conjunction with the second-order Trotter formula. The populations obtained deviate strongly from the reference, which is due to the error of the Trotterization within a time step. Only with a smaller time step size of $0.2$ a.u. (purple and red dashed lines), convergence is reached for the quantum simulations and the populations do not differ from the reference by more than $10^{-2}$ at any time step. For higher Trotter orders, larger time steps would also be sufficient, but lead to longer computation times.

Besides state populations, expectation values of various operators are often of interest for the description of the electron dynamics. For example is the time-dependent dipole moment of the molecule that reflects the transient displacement of electron density (negative partial charge) compared to the nuclear positions (positive charge). The reference traditional TD-CI dipole moment of the hydrogen molecule during the state-to-state transition was determined at each time step. It is shown as a red line in Fig. \ref{fig:h2} (b). As expected, it oscillates with the driving laser field as discussed in more detail by Krause et al.\cite{Krause2007} 

The purple line in panel (b) shows the respective dipole moment obtained by the quantum computing algorithm. Here, the time-dependent dipole moment is determined after each time step using the Hadamard test (Sec. \ref{SecTheo_QComp_Prop}). For this purpose, the dynamics is repeatedly simulated up to the corresponding time step and the Hadamard test with a $\Delta x = 0.5$ is then applied 20000 times. The comparison to the TD-CI graph shows that the dipole moment can be satisfactorily reproduced apart of introduction of some numerical noise. The slight deviations are connected with the number of repetitions of the Hadamard test. The noise is reduced with increasing number of repetitions. Thus, we expect improvement and convergence to the exact TD-CI result for an even larger number of repetitions.

\subsection{Charge-transfer Dynamics in LiH}\label{SecRes_LiH}
For LiH also simulations for a state-to-state transition by $\pi$ pulses are performed (cf. \ref{SecComp}). Here, the target state signifies are charge transfer, e.i., the change in the dipole moment along the bond axis reflects the transfer of an electron form the hydrogen atom to the lithium atom.

\begin{figure}
\includegraphics[width=0.48\textwidth,clip=]{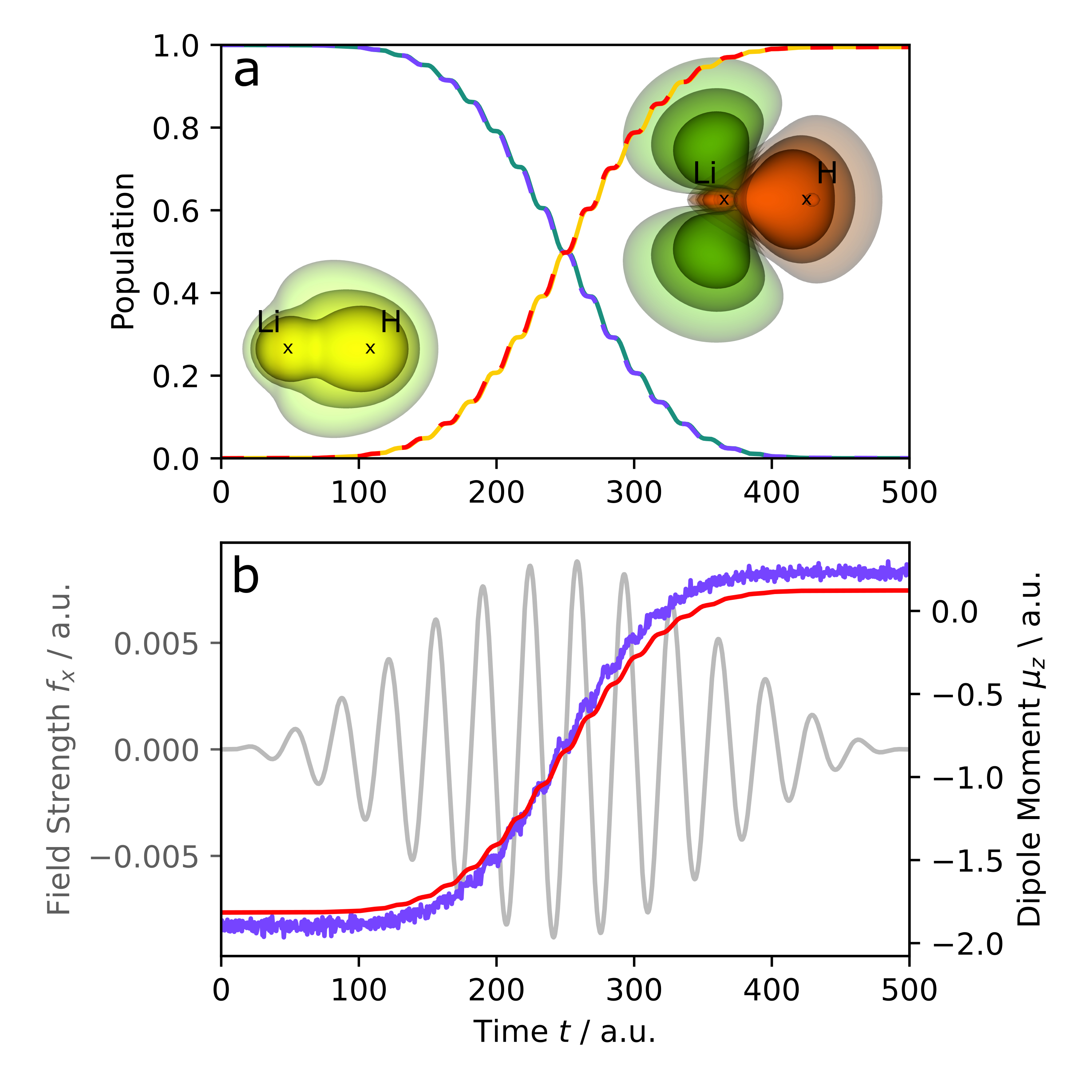}
\caption{\label{fig:lihlong} 
Comparison of state population (a) and dipole moment (b) with TD-FCI ((a) solid, (b) red) and quantum computer dynamics ((a) dashed, (b) purple) of charge transfer in lithium hydride during a resonant \textit{weak} $\pi$ pulse excitation ((b) gray), as well as the electron density at the beginning and the difference density at the end.
}
\end{figure}

The populations for LiH states upon resonant laser excitation are shown in Fig. \ref{fig:lihlong} (a) again as yellow and turquoise solid lines for the TD-CI calculation and for the quantum computer calculation as red and purple dashed lines. Again, the initial ground state density is shown on the left and the difference density with respect to the initial state is shown on the right. The $x$-polarized $\pi$ pulse can be seen in panel (b) as gray line together with the time-dependent dipole moment along the $z$ direction as obtained for standard TD-CI (red) and from the quantum simulation (purple). 

The physical interpretation of panel (a) is straightforward: A population inversion from the ground into the second excited singlet state occurs. After the laser pulse, no further dynamics can be detected as propagation proceeds (not shown). What is different compared to the H$_2$ example is that the final state accumulates density centered at the Li atom in form of a $\pi_x$ orbital perpendicular to the bonding axis, as seen in the positive difference density (green). This combines with a continuous change of the electronic dipole moment. Both observations signify the process of charge-transfer along the bond axis. However, since our simulations for numerical algorithms employ a minimal basis set, the experimental values are not reproduced. 

The populations of the Hamiltonian simulations are congruent to the TD-CI calculation, when a Trotter decomposition of second order in conjunctions with a time step of $0.2$ a.u. is applied. The evolution of the dipole moment obtained from the Hadamard test with a $\Delta x = 0.2$ and with 20000 repetitions shows both a noisy shape and a small deviation from TD-CI results, especially at the beginning and end of the dynamics. Again with more repetitions and Trotter parameters specifically optimized for the Hadamard test, an improvement to the already accurate results could be achieved.

\begin{figure}
\includegraphics[width=0.48\textwidth,clip=]{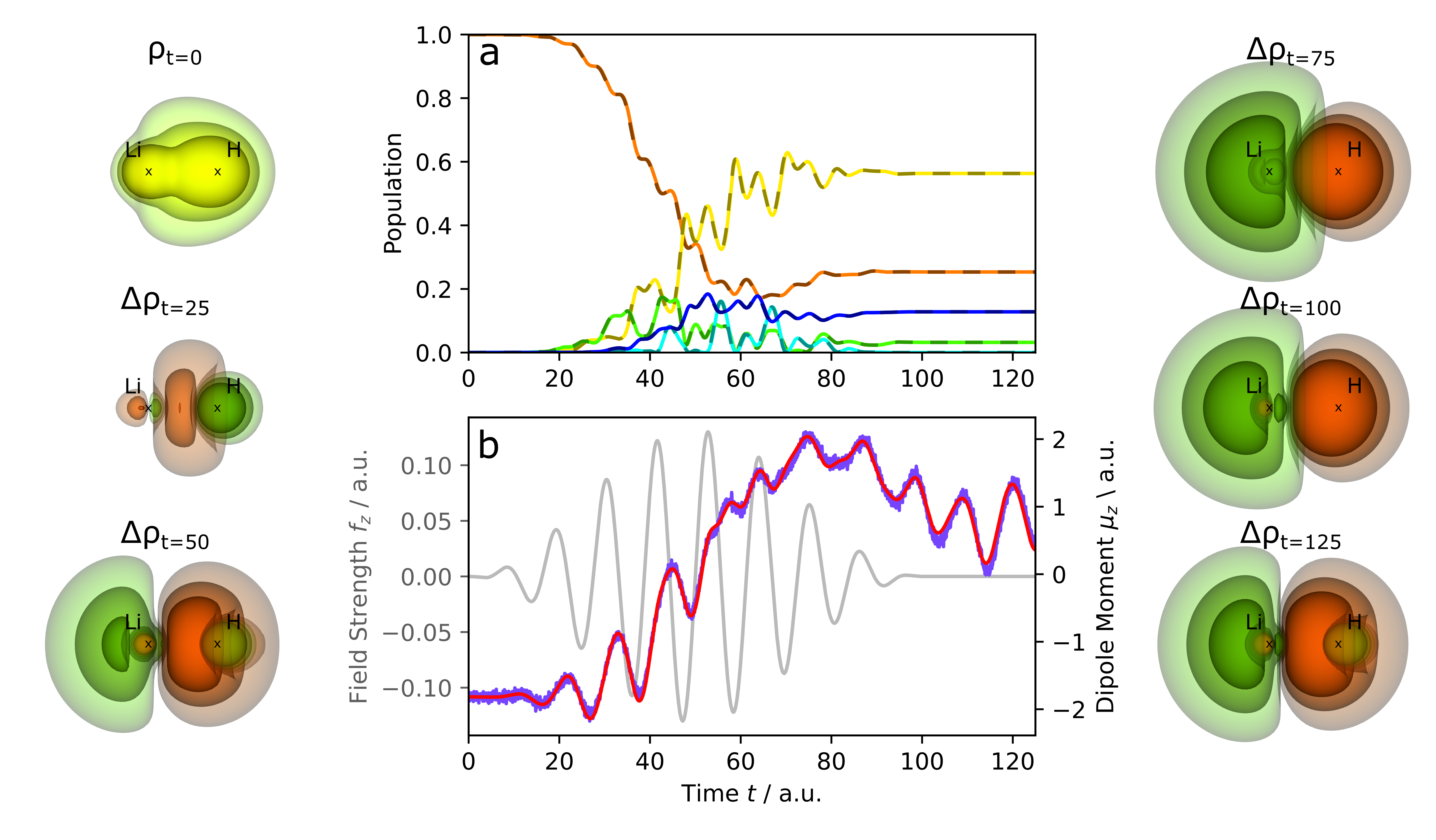}
\caption{\label{fig:lihshort}Comparison of state population (a) and dipole moment (b) with TD-FCI ((a) solid, (b) red) and quantum computer dynamics ((a) dashed, (b) purple) of charge transfer in lithium hydride during an \textit{intense} $\pi$ pulse excitation ((b) gray), as well as the electron density at the beginning and the difference density between initial density and electron density at different times.}
\end{figure}

As a second calculation on the lithium hydride molecule, the dynamics induced by a short and intense laser pulse is studied. In this example, the frequency of this $\pi$ pulse is chosen resonant to the fifth ecxcited state. The central panels (a) and (b) of Fig. \ref{fig:lihshort} are composed in the same way as for the previous examples. Inspecting the TD-CI result first (solid lines in (a), red line in (b), it is obvious, that the dynamics is much richer than in the previously-discussed weak, long pulse excitation, because the strong $z$-polarized pulse (Fig. \ref{fig:lihshort}(b), gray) creates a wave packet. This means that many different states are excited simultaneously or sequentially with the targeted, resonant fifths excited state. Some of the states have no significant population, but there are five states involved, which have at least 1 \% overlap with the total wavefunction. They are shown with solid lines for the TD-CI calculation in panel (a), where respective dashed lines represent the result of the quantum computation. The state populations are fluctuating significantly during intense cycles of the pulse, but stabilize thereafter. The $z$-component of the dipole moment expectation value (b) likewise fluctuates, reflecting rich dynamics of the electron density. Such oscillations continue even after the laser pulse is off, signifying the creation of a wave packet. Tracking the difference densities $\Delta \rho$ at different times, a charge-transfer process from hydrogen to lithium can be seen as well, however, in this case, mostly a $\sigma^*$ orbital extending beyond the Li atom was addressed. This causes the dipole moment along the bond to change more drastically by about $4$ a.u. compared to $2$ a.u. previously. A second distinction is that the difference density very close to the atomic centers oscillates. This correlates to the dipole moment such that its local maxima fall together with difference density accumulation (green) and the minima with density decrease (red) on the H atom. Finally, even after the laser is no longer acting after $t=100$ a.u. and populations are no longer changing, the electron density, and thus also dipole moment expectation value, do not stabilize.

Turning finally to the comparison with the quantum computation of this clearly more complicated example, in particular the populations (a) of TD-CI and Hamilton simulation (dashed lines) agree to the third decimal point with using a second-order Trotter decomposition and a time step of $0.1$ a.u. The dipole moment obtained from 50000 Hadamard repetitions and $\Delta x = 0.1$ (blue line in (b)) also agrees very well apart from the noise. This leads to the conclusion that even for wave packet dynamics, quantum simulation and Hadamard test return reliable numerics for the electron dynamics of a fully-correlated wavefunction (FCI).

\subsection{Ionization Dynamics of H$_2$}\label{SecRes_H2Ion}
This chapter comprises results on the extension of quantum compute tools for electron dynamics towards the delicate problem of non-Hermitian operators in the time evolution. This exemplified by addition of a CAP operator allowing to account for an ionization process of the H$_2$ molecule. A resonant $z$-polarized laser excites H$_2$ into a state that lies in the continuum. Since our simulations employ a minimal basis, the first excited state has an energy above the ionization potential and can be used for the heuristic model as described before. Therefore, the same laser pulse is applied as in Sec. \ref{SecRes_H2St2St} for the numerical determination for the action of a non-Hermitian operator.

\begin{figure}
\includegraphics[width=0.48\textwidth,clip=]{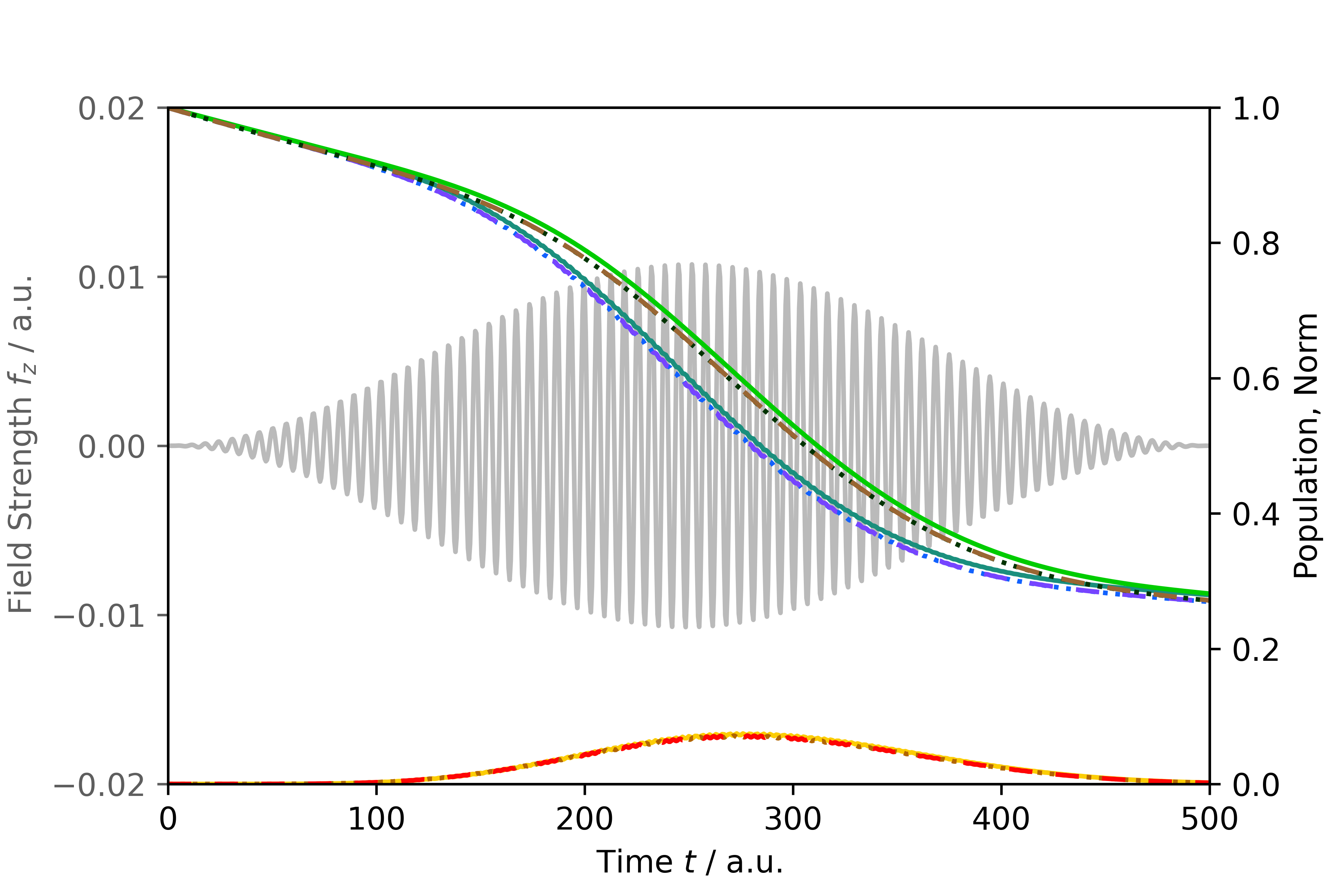}
\caption{\label{fig:h2cap}Comparison of the populations and norm of a TD-CI calculation (solid, populations turquoise and yellow, norm light green) and a quantum computer calculation with QITE algorithm once with exact expected values (dotted, populations blue and orange, norm dark green) and once with measured expected values (dashed, populations purple and red, norm brown) during an ionization dynamic with a resonant laser excitation (gray) in a hydrogen molecule.}
\end{figure}

Fig. \ref{fig:h2cap} shows the populations and the norm during the ionization dynamics. The solid lines belong to a TD-CI calculation, where turquoise (yellow) represents the ground (excited) state population and light green the norm. The time-dependent field polarized along the bond axis is shown in gray.

During excitation the population is transferred from the ground state into the excited state by the laser pulse. However, since the excited state is located in the ionization continuum, it is immediately depopulated through electron density removal by the CAP. This can also be seen in the norm loss. With increasing degree of excitation towards the center of the laser pulse, not only the population of the excited state but also the norm loss increases. With subsequent weakening of the pulse, the population of the excited state decreases and causes a further loss of norm which terminates when the excited state's population reaches zero. The remaining population in the ground state becomes constant.

For the quantum computer dynamics, the expectation values from the QITE algorithm are at first read out exactly and directly from the quantum computer simulator. The resulting ground (excited) state population is shown with dotted blue (orange) lines and the norm is shown as dotted dark green line. In an alternative quantum computer simulation, the expectation values are determined by $10^{6}$ measurements each. The ground (excited) state population thereof is depicted as dashed purple (red) line and the norm as dashed brown line. In both readout cases the ground state population is in perfect agreement with the TD-CI reference, whereas the excited state population and the norm are slightly underestimated. However, they agree exactly among measured and exact readout in the QITE algorithm. Thus, ionization dynamics with non-Hermitian operators and measured expectation values could be executed and fully analyzed on real quantum computers.

\section{Discussion}\label{SecDiscussion}
While the exact time-dependent full configuration interaction method, used here for small molecules, can never be done for larger chemical structures due to exponential scaling, this work demonstrate how quantum computer algorithms can be employed to overcome this limit. Each of the algorithms introduced may scale polynomially: the Trotter decomposition scales as $O(poly(1/\epsilon))$ depending on encoding, propagation algorithms, desired accuracy $\epsilon$, and molecule. For common molecules, for example, a scaling of $O(N^6)$ with the number of spin orbitals $N$ was shown.\cite{Poulin2014}
Beyond this, the scaling of the Hadamard test for the determination of expectation values depends additionally on the desired accuracy $\epsilon$ with $O(1/\epsilon)$. The major challenge with respect to scaling lies in the dynamics with non-Hermitian operators for which this study offers one solution. Since the Jordan-Wigner encoding stores the parity of a state on the entire qubit register, all qubits are correlated with each other. This leads in the QITE algorithm to a exponentially increasing number of expectation values with the qubits, which makes an application on molecules with more spin orbitals not possible. However, with a different encoding and further simplifications, the algorithm could be improved in terms of scaling other than a propagation on a classical computer.

\section{Conclusion}\label{SecConclusion}
In this work, we demonstrated how algorithms for error-free quantum computing can be applied to enable electron dynamics simulations. First, the hydrogen molecule in its minimal basis was propagated under the effect of a $\pi$ pulse as a most simple test system to compare the established time-dependent full configuration interaction method for classical computers with our quantum-computer algorithms. On the quantum-computer algorithm side, the Jordan-Wigner encoding translates the fermionic initial state and propagator to qubits. For the propagation, the Trotter decomposition served the expression of quantum gate sequences. Finally, a comparison of the state populations of the classical and quantum computer algorithms showed under which conditions the quantum propagation reproduces the TD-CI one. Likewise in hydrogen, it was then proven that propagations including non-Hermitian operators, such as complex absorbing potentials for capturing ionization, are also possible. For this purpose, a QITE-like algorithm was applied.

For analysis of electron dynamics the time evolution of expectation values is frequently considered. For achieving this on a quantum computer, the Hadamard test was applied. Computations of the dipole moment of the LiH molecule in two different scenarios, a long $\pi$ as well as a short intense pulse demonstrate the applicability of the methods to significantly more complex processes with several simultaneously excited state populations. 

Electron dynamics in atomic and molecular systems is an area of research where experimental techniques are developed quickly. They enable the study of processes that can only be simulated with high-precision quantum-dynamics methods. On current conventional computers this is possible only for a very limited extent. Quantum computing, with the algorithms shown, may bring a breakthrough in this field by enabling calculations on reasonably large molecules that are otherwise impossible. This advance in quantum simulations transfers with the same merits to the field of nuclear quantum dynamics.

\begin{acknowledgments}
The authors are grateful to the financial support provided from Volkswagen Foundation via the Freigeist Fellowship (No. 89525). They thank Dr. Pascal Krause for discussions on the TD-CI technique in conjunctions with complex-absorbing potentials as well for a first revision of the manuscript. Finally, they acknowledge the individualized and knowledgeable setup of the Helmholtz-Zentrum Berlin's compute server for quantum compute simulations done by Robert Grzimek.
\end{acknowledgments}

%\section*{AIP Publishing Data Sharing Policy}
%The data that support the findings of this study are available from the corresponding author upon reasonable request.

%\bibliographystyle{apsrev4-2}
\bibliography{references}
\end{document}